\documentstyle[twoside,fleqn,espcrc2,epsfig]{article}

\def\rmO{{\rm O}}

\def\proof{\noindent{\sl Proof:}\kern0.6em}

\def\frac#1#2{\hbox{$#1\over#2$}}
\def\dual{\mathstrut^*\kern-0.1em}

\def\lvec#1{\setbox0=\hbox{$#1$}
    \setbox1=\hbox{$\scriptstyle\leftarrow$}
    #1\kern-\wd0\smash{
    \raise\ht0\hbox{$\raise1pt\hbox{$\scriptstyle\leftarrow$}$}}
    \kern-\wd1\kern\wd0}
\def\rvec#1{\setbox0=\hbox{$#1$}
    \setbox1=\hbox{$\scriptstyle\rightarrow$}
    #1\kern-\wd0\smash{
    \raise\ht0\hbox{$\raise1pt\hbox{$\scriptstyle\rightarrow$}$}}
    \kern-\wd1\kern\wd0}

\def\nabstar#1{\nabla\kern-0.5pt\smash{\raise 4.5pt\hbox{$\ast$}}
               \kern-4.5pt_{#1}}

\def\drvstar#1{\partial\kern-0.5pt\smash{\raise 4.5pt\hbox{$\ast$}}
               \kern-5.0pt_{#1}}

\def\Nf{N_{\rm f}}

\def\rhoprime{\rho\kern1pt'}
\def\rhobar{\bar{\rho}}
\def\rhobarprime{\rhobar\kern1pt'}
\def\rhobartilde{\kern2pt\tilde{\kern-2pt\rhobar}}
\def\rhobartildeprime{\kern2pt\tilde{\kern-2pt\rhobar}\kern1pt'}

\def\zetabar{\bar{\zeta}}
\def\zetaprime{\zeta\kern1pt'}
\def\zetabarprime{\zetabar\kern1pt'}
\def\zetar{\zeta_{\raise-1pt\hbox{\sixrm R}}}
\def\zetabarr{\zetabar_{\raise-1pt\hbox{\sixrm R}}}

\def\phiimpr{\phi_{\kern0.5pt\hbox{\sixrm I}}}

\def\diracstar#1#2{
    \setbox0=\hbox{$\gamma$}\setbox1=\hbox{$\gamma_{#1}$}
    \gamma_{#1}\kern-\wd1\kern\wd0
    \smash{\raise4.5pt\hbox{$\scriptstyle#2$}}}

\def\bv{b_{\rm V}}

\def\bs{b_{\rm S}}

\def\bx{b_{\rm X}}

\def\bg{b_{\rm g}}
\def\bm{b_{\rm m}}

\def\ca{c_{\rm A}}
\def\cv{c_{\rm V}}
\def\cT{c_{\rm T}}

\def\fv{f_{\rm V}}
\def\f1{f_1}

\def\CF{C_{\rm F}}

\def\opprime#1{\setbox0=\hbox{${\cal O}$}\setbox1=\hbox{${\cal O}_{\rm #1}$}
    {\cal O}_{\rm #1}\kern-\wd1\kern\wd0
    \smash{\raise4.5pt\hbox{\kern1pt$\scriptstyle\prime$}}\kern1pt}

\def\ophatprime#1{\setbox0=\hbox{$\widehat{\cal O}$}
    \setbox1=\hbox{$\widehat{\cal O}_{\rm #1}$}
    \widehat{\cal O}_{\rm #1}\kern-\wd1\kern\wd0
    \smash{\raise4.5pt\hbox{\kern1pt$\scriptstyle\prime$}}\kern1pt}

\def\bopprime#1{\setbox0=\hbox{${\cal O}$}\setbox1=\hbox{${\cal O}_{\rm #1}$}
    {\cal L}_{\rm #1}\kern-\wd1\kern\wd0
    \smash{\raise4.5pt\hbox{\kern1pt$\scriptstyle\prime$}}\kern1pt}

\def\blagprime#1{\setbox0=\hbox{${\cal B}$}\setbox1=\hbox{${\cal B}_{#1}$}
    {\cal B}_{#1}\kern-\wd1\kern\wd0
    \smash{\raise5.2pt\hbox{\kern1pt$\scriptstyle\prime$}}\kern1pt}

\def\gr{g_{{\hbox{\sixrm R}}}}

\def\mq{m_{\rm q}}

\def\mc{m_{\rm c}}

\def\zm{Z_{\rm m}}

\def\msbar{{\rm \overline{MS\kern-0.05em}\kern0.05em}}

\title{Further one-loop results in O($a$) improved lattice 
QCD\thanks{talk given by S.~Sint at the International Symposium on 
Lattice Field Theory, July 22--26, 1997, Edinburgh}}

\author{
        Stefan Sint\address{SCRI, The Florida State University, 
        Tallahassee, Florida 32306--4130}
and
        Peter Weisz\address{Max-Planck-Institut f\"ur Physik, F\"ohringer 
                      Ring 6, D-80805 M\"unchen, Germany}}

\begin{document}
\begin{abstract}
Using the Schr\"odinger functional we have computed
a variety of renormalized on-shell correlation functions to one-loop 
order of perturbation theory. By studying their approach
to the continuum limit we have determined the O($a$) counterterms needed
to improve the quark mass and a number of isovector quark bilinear 
operators.

%To this order we then also obtain
%the relation between the renormalized quark mass in the minimal
%scheme of dimensional regularization and 
%a renormalized quark mass derived from the Schr\"odinger functional.
\end{abstract}
\maketitle

\section{INTRODUCTION}

Recent work by the ALPHA collaboration has focused on 
non-perturbative renormalization and on-shell
O($a$) improvement of lattice QCD with Wilson quarks~[1--7].
Various improvement coefficients could be determined as functions
of the bare coupling $g_0$, including the coefficient
of the Sheikholeslami-Wohlert term in the lattice action~\cite{SW}.

While a non-perturbative determination of the improvement coefficients
is clearly preferable perturbative estimates are 
nevertheless useful. First of all, any non-perturbative
determination should establish contact with perturbation theory
at sufficiently small values of the bare coupling constant. 
This provides a non-trivial check for the chosen strategy
and criteria to assess the quality of a given 
improvement condition. Secondly, for those coefficients 
which account for lattice effects due
to non-zero quark masses, perturbative estimates may indeed 
be satisfactory provided the quark masses are small
when measured in lattice units.

In this contribution we present our one-loop results
for the on-shell O($a$) improved isovector 
composite operators which are bilinear in the quark fields.
The computational strategy and most of the results 
have already been published in refs.~\cite{paperII,paperV}. 
In addition we here also include the results for 
the improved isovector tensor and scalar densities.

\section{DEFINITIONS}

We consider lattice QCD with
$\Nf\ge 2$ degenerate quark flavours of bare mass $m_0$ 
and shall assume that the action has already been on-shell 
improved~\cite{SW}.
We are interested in the improvement of the 
following isovector operators,
\begin{eqnarray}
  V^a_{\mu}(x)&=&\bar{\psi}(x) \gamma_{\mu}\frac12\tau^a\psi (x),\\
  A^a_{\mu}(x)&=&\bar{\psi}(x) \gamma_{\mu}\gamma_5\frac12\tau^a\psi (x),\\
  P^a (x)&=&\bar{\psi}(x) \gamma_5\frac12\tau^a\psi (x),\\
  S^a (x)&=&\bar{\psi}(x) \frac12\tau^a\psi (x),\\
  T^a_{\mu\nu}(x)&=&i\bar{\psi}(x) \sigma_{\mu\nu}\frac12\tau^a\psi (x).
\end{eqnarray}
Here $\tau^a$ ($a=1,2,3$) are the usual Pauli matrices acting in any
two-flavour subspace and our conventions for the Dirac matrices
are as in ref.~\cite{paperI}.

In a mass independent renormalization scheme 
the renormalized (at renormalization scale $\mu$)
and O($a$) improved counterparts of the 
above fields all take the form ($X=V,A,P,S,T$)~\cite{paperI}
\begin{equation}
   X_{\rm R} = Z_{\rm X}(\tilde{g}_0^2,a\mu)
            \bigl[1+b_{\rm X}(g_0^2) a\mq\bigr]X_{\rm I},
 \label{normalisation}
\end{equation}
where  $X_{\rm I}$ stands for
\begin{eqnarray}
  (V_{\rm I})^a_{\mu}&=&V^a_{\mu}+\cv(g_0^2) a
  \tilde\partial_{\nu}T^a_{\mu\nu},\\
  (A_{\rm I})^a_{\mu}&=&A^a_{\mu}+\ca(g_0^2) a
  \tilde\partial_{\mu}P^a,\\
  (T_{\rm I})^a_{\mu\nu}&=&T^a_{\mu\nu}+\cT(g_0^2) a
  (\tilde\partial_{\mu}V^a_{\nu}-\tilde\partial_{\nu}V^a_{\mu}),
\end{eqnarray}
and otherwise $X_{\rm I}=X$.
Here $\tilde\partial_{\mu}$ denotes the symmetric lattice derivative
and the parameter $\tilde{g}_0$ is connected to the bare coupling
$g_0$ through
\begin{equation}
  \tilde{g}_0^2 =g_0^2 \bigl[ 1+\bg(g_0^2)a\mq \bigr],
\end{equation}
where $\mq=m_0-\mc$ and $\mc$ is the critical bare quark mass.
Similarly one defines
\begin{equation}
  \tilde{m}_{\rm q}=\mq\bigl[ 1+\bm(g_0^2)a\mq \bigr],
\end{equation}
and the renormalized O($a$) improved
coupling and quark mass 
are then related to these parameters by~\cite{paperI},
\begin{eqnarray}
 \gr^2     &=&\tilde{g}_0^2Z_{\rm g}(\tilde{g}_0^2,a\mu),\\
  m_{\rm R}&=&\tilde{m}_{\rm q}\zm(\tilde{g}_0^2,a\mu).
\end{eqnarray}
The improvement coefficients 
$\bg,\bm,\bx$  can be expanded in perturbation theory,
\begin{equation}
  b=b^{(0)} + b^{(1)}g_0^2 +\rmO(g_0^4),
\end{equation}
and an analogous expansion exists for
the coefficients $\ca,\cv$ and~$\cT$.  

\section{COMPUTATIONAL STRATEGY}

To determine the improvement coefficients we chose to 
compute a number of on-shell correlation functions
derived from the Schr\"odinger functional~(SF).
The SF is the Euclidean functional integral for QCD on
a finite space-time manifold where the (spatially periodic)
quantum fields satisfy Dirichlet boundary conditions 
in the time direction~\cite{alphaI,StefanI}.
For proper choice of the boundary conditions it can be shown that
the lattice action has a unique absolute minimum~\cite{alphaI}. 
The saddle point expansion about this minimum is then straightforward
(albeit technically involved), and zero modes do not appear.

The gauge field boundary conditions imply that 
only global gauge transformations are allowed at the boundaries.
Therefore, gauge invariant correlation functions can be defined 
where the quark and antiquark fields 
at the boundaries are separately projected onto their zero spatial
momentum components. This is convenient because the 
perturbative expansion of such a correlation function 
starts with tree diagrams. Furthermore, 
exactly the same correlation functions
can be used in numerical simulations.

In order to take the continuum limit in a finite space-time
volume one fixes the time extent $T$, the renormalized O($a$)
improved quark mass and all other dimensionful parameter 
in units of $L$, the spatial extent of the space-time manifold.
As a result O($a$) lattice artefacts always appear as 
$a/L$ effects and can be identified by varying 
the lattice size. In each renormalized correlation function these
effects are cancelled by an a priori different
linear combination of O($a$) improvement coefficients.
The finite volume provides a great flexibility here because many
different renormalized correlation functions can be obtained
by simple changes of the boundary conditions.

For completeness we mention that 
Dirichlet boundary conditions cause 
additional divergences and O($a$) artefacts
localised at the boundaries.
These can be absorbed by renormalizing the boundary
quark and antiquark fields in the same way
as the composite fields in eq.~(\ref{normalisation}),
and by including additional O($a$) boundary 
counterterms in the lattice action~\cite{alphaI,StefanI,paperI}. 

For details of our computational strategy and the definitions of
most of the correlation functions the reader should 
consult refs.~\cite{paperII,paperV}. 
We have treated the case of the isovector tensor density 
in complete analogy to the improved vector current, and 
the computation of $\bs$ involved a correlation function
similar to $\fv$ of ref.~\cite{paperIV},
where the vector current was replaced by the isovector scalar density. 

\begin{table}[h]
%\centering
\vskip -2ex
\begin{tabular} {| c | c | r |c|}
  \hline 
 $\phantom{01234}$&$\phantom{0123456}$&$\phantom{0123456}$
 &$\phantom{0123}$\\[-2ex]
 ${\rm X}$ & $b_{\rm X}^{(0)}$ &
            \multicolumn{1}{c|}{$b_{\rm X}^{(1)}$}& ref.\\[1ex]
 \hline
  & & &\\[-2ex]
     ${\rm g}$ 
  &  $0$ 
  &  $0.012000(2)\times\Nf$ 
  &\cite{StefanRainer}\\[1ex]
     ${\rm m}$ 
  &  $-\frac12$ 
  &  $ -0.07217(2)\times\CF$ 
  &\cite{paperV} \\[1ex]
%%
%     $\zeta$ 
%  &  $-\frac12$ 
%  &  $ -0.06738(4)\times\CF$ 
%  &\cite{paperV} \\[1ex]
%%
     ${\rm V}$                              
  &  $1$ 
  &  $\phantom{-}0.11492(4)\times\CF$
  &\cite{paperV} \\[1ex]
     ${\rm A}$ 
  &  $1$ 
  &  $\phantom{-}0.11414(4)\times\CF$
  &\cite{paperV} \\[1ex]

     ${\rm P}$                              
  &  $1$ 
  &  $\phantom{-}0.11484(2)\times\CF$
  &\cite{paperV} \\[1ex]  
     ${\rm S}$                              
  &  $1$ 
  &  $\phantom{-}0.14434(5)\times\CF$
  &\\[1ex]                
     ${\rm T}$                              
  &  $1$ 
  &  $\phantom{-}0.10434(4)\times\CF$
  &\\[1ex]                 
\hline
\end{tabular}
\vskip 2ex
{Table 1. Improvement coefficients $b$}
\label{b_coeff}
\vskip -4ex
\end{table}

\section{RESULTS}

To one-loop order of perturbation theory we have carried out
many consistency checks and thus confirm the general 
framework of O($a$) improvement as described in ref.~\cite{paperI}.

Numerically we obtain ($\CF=(N^2-1)/2N$ for $N$ colours),
\begin{eqnarray}
 \ca &=& -0.005680(2)\times \CF g_0^2+ \rmO(g_0^4),\\ 
 \cv &=& -0.01225(1)\times \CF g_0^2+ \rmO(g_0^4),\\
 \cT &=& \phantom{-}0.00896(1)\times \CF g_0^2+ \rmO(g_0^4).
\end{eqnarray}
The coefficient $\ca$ has first been obtained in ref.~\cite{paperII}.
and  $\cv$ was given in ref.~\cite{paperV}. 
The results and references for the $b$-coefficients 
are collected in table~1.
Note that to order $g_0^2$ we find, within errors,
\begin{equation}
 \bs=-2\bm. 
 \label{bsm}
\end{equation}
In fact, it can be shown that eq.~(\ref{bsm}) 
is an exact identity in quenched QCD and furthermore
the isoscalar scalar operator has the same $b$-coefficient 
as the isovector~\cite{Martin}.  

Comparison with the non-perturbative results
for $\ca$~\cite{paperIII} and $\cv$~\cite{MarcoRainer} 
shows that in these cases 
perturbative estimates are not accurate at large 
values of the bare coupling constant.
In the case of $\bv$ we can compare with the 
non-perturbative result of ref.~\cite{paperIV}. 
In figure~1 we see that the non-perturbative 
values (represented by the solid line) are quite
a bit higher than the perturbative estimate (dotted line), even when
Parisi's boosted coupling~\cite{Parisi} is used (crosses).
However, using the boosted perturbative one-loop estimate  at $g_0=1$
an error of only 1 per cent is induced in the normalisation factor 
$1+\bv a\mq$~[cf.~eq.~(\ref{normalisation})], 
provided $a\mq\leq 0.05$. 

%If we are willing to tolerate a maximal relative error 
%of 1 (3) per cent in the normalisation factor $1+\bv a\mq$
%the values for $a\mq$ must not exceed $0.02$ ($0.07$), $0.03$ ($0.11$)
%and $0.05$ ($0.17$) for the tree level, one-loop 
%and boosted one-loop estimates of $\bv$ respectively. 

It is of course not clear whether similar conclusions can be drawn
for the other $b$-coefficients.
Further non-perturbative results would obviously 
be welcome and some progress in this direction
has been reported at this conference~\cite{GiuliaRoberto,Sharpe}.

\begin{figure}[t]
%\vskip -2.5ex
\epsfig{file=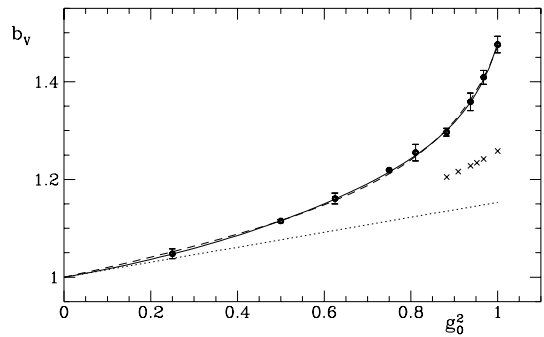,%18 144 592 718
%       bbllx=68pt,%
%       bblly=144pt,%
%       bburx=642pt,%
%       bbury=768pt,%
%        clip=,%
       bbllx=68pt,%
       bblly=250pt,%
       bburx=642pt,%
       bbury=568pt,%
        clip=,%
       width=8cm}%
%      height=10cm}
%\framebox[55mm]{\rule[-21mm]{0mm}{43mm}}
\vskip -5ex
\caption{Improvement coefficient $\bv$ for gauge group SU($3$)
and $\Nf=0$}
\vskip -4ex
\label{fig:largenenough}
\end{figure}

\vskip 1ex

This work is part of the ALPHA collaboration research programme.
We would like to thank M.~L\"uscher, S.~Sharpe and U.~Nierste
for discussions. 
S.~Sint acknowledges support by the U.S. Department of Energy
(contracts DE-FG05-85ER250000 and DE-FG05-96ER40979).

% List of references

\end{document}